\documentclass[twocolumn,trackchanges]{aastex6}
\RequirePackage{lineno}
\usepackage{amsmath}
\usepackage{amssymb}
\usepackage{amsthm}
\usepackage{natbib,aasdefs,url,bm}
\usepackage{array}
\usepackage{float}
\usepackage{lipsum,graphicx}
\usepackage{subfigure}
\usepackage{color}
\def\mr#1{\mathrm{#1}}

\newcounter{ichi}
\newcounter{ni}
\newcounter{san}
\newcounter{yon}
\def\be{\begin{equation}}
\def\ee{\end{equation}}
\def\ba{\begin{eqnarray}}
\def\ea{\end{eqnarray}}

\def\aap{\ {A\&A}\ }

\def\apjl{\ {ApJL}\ }
\def\apjs{\ {ApJS}\ }

\shorttitle{Revealing the Production Mechanism of High-Energy Neutrinos from NGC 1068}
\shortauthors{Das, Zhang \& Murase}

\begin{document}

\title{Revealing the Production Mechanism of High-Energy Neutrinos from NGC 1068}

\author{Abhishek Das\altaffilmark{1,2,3}}

\author{B. Theodore Zhang\altaffilmark{4}}

\author{Kohta Murase\altaffilmark{1,2,3,4}}
\altaffiltext{1}{Department of Physics, The Pennsylvania State University, University Park, PA 16802, USA}
\altaffiltext{2}{Department of Astronomy \& Astrophysics, The Pennsylvania State University, University Park, PA 16802, USA}
\altaffiltext{3}{Center for Multimessenger Astrophysics, Institute for Gravitation and the Cosmos, The Pennsylvania State University, University Park, PA 16802, USA}
\altaffiltext{4}{Center for Gravitational Physics and Quantum Information, Yukawa Institute for Theoretical Physics, Kyoto University, Kyoto 606-8502, Japan}

\begin{abstract}
The detection of high-energy neutrino signals from the nearby Seyfert galaxy NGC 1068 provides us with an opportunity to study nonthermal processes near the center of supermassive black holes. Using the IceCube and latest {\it Fermi}-LAT data, we present general multimessenger constraints on the energetics of cosmic rays and the size of neutrino emission regions. 
In the photohadronic scenario, the required cosmic-ray luminosity should be larger than $\sim1-10$\% of the Eddington luminosity, and the emission radius should be $\lesssim15~R_S$ in low-$\beta$ plasma and $\lesssim3~R_S$ in high-$\beta$ plasma. The leptonic scenario overshoots the {\it NuSTAR} or {\it Fermi}-LAT data for any emission radii we consider, and the required gamma-ray luminosity is much larger than the Eddington luminosity. The beta decay scenario also violates not only the energetics requirement but also gamma-ray constraints especially when the Bethe-Heitler and photomeson production processes are consistently considered. Our results rule out the leptonic and beta decay scenarios in a nearly model-independent manner, and support hadronic mechanisms in magnetically-powered coronae if NGC 1068 is a source of high-energy neutrinos.  
\end{abstract}


\section{Introduction}
The sources of high-energy cosmic neutrinos have remained a big mystery since the discovery by the IceCube Collaboration~\citep{Aartsen:2013bka, Aartsen:2013jdh}. 
Inelastic hadronuclear ($pp$) and/or photohadronic ($p\gamma$) processes that generate neutrinos should also produce gamma rays of similar energies. 
The multimessenger observations of the all-sky neutrino and gamma-ray fluxes led to the conclusion that the dominant sources of $10-100$~TeV neutrinos in the range are hidden or opaque for GeV--TeV gamma rays~\citep{Murase:2013rfa,Murase:2015xka, Capanema:2020rjj, Fang:2022trf}.
Recently, the IceCube collaboration reported an excess of 79 events with a global significance of 4.2$\sigma$~\citep{IceCube:2022der}.
These neutrinos are observed in the $\sim$1.5 to $\sim$15 TeV energy range, which leads to a neutrino luminosity about one or two orders of magnitude larger than the gamma-ray luminosity in the GeV--TeV energy range~\citep{Fermi-LAT:2019pir, Fermi-LAT:2019yla, MAGIC:2019fvw, Ajello:2023hkh}. 
This implies that the source must be opaque to gamma rays, which supports the results from the diffuse background observations. Therefore, NGC 1068 is treated as a hidden neutrino source, and particle acceleration and secondary emission are very likely in the inner disk and corona regions. The multimessenger data constrain the size of the emission region to $\lesssim 30-100 R_S$, where $R_S$ is the Schwarzschild radius \citep{Murase:2022dog}.
The population of jet-quiet active galactic nuclei (AGNs) can explain the all-sky neutrino flux particularly in the 10-100~TeV range~\citep{PhysRevLett.125.011101,Eichmann:2022lxh, Padovani:2024tgx}, and higher energies may be explained by jet-loud AGNs~\citep{Fang:2017zjf} or low-luminosity AGNs~\citep{Kimura:2014jba, Kimura:2020thg}.

The mechanisms of high-energy neutrino emission and particle acceleration in the vicinity of supermassive black holes (SMBHs) have been under debate~\citep[see a review][and references therein]{Murase:2022feu}. It is believed that Seyfert galaxies and quasars commonly have a big blue bump that is attributed to multitemperature black body emission from an accretion disk, as well as power-law X-ray emission from a hot plasma region called a corona. Some mechanisms rely on magnetically-powered corona models, where particles are accelerated by turbulence and/or magnetic reconnections in low-$\beta$ plasma~\citep{PhysRevLett.125.011101,Kheirandish:2021wkm, Eichmann:2022lxh,Fiorillo:2023dts, Mbarek:2023yeq}. Other mechanisms involve shocks, which could be caused by the free fall of the material~\citep{Inoue:2019yfs}, failed winds~\citep{Inoue:2022yak}, and reconnection-driven flows~\citep{Murase:2022dog}, where the plasma is not magnetically dominated. High-energy neutrinos are presumably produced via hadronic processes, but in principle they could also be produced by leptonic processes \citep[e.g.,][]{Bhattacharjee_2000, Athar_2001, li2007eev,Hooper:2023ssc} and neutron decay from the photodisintegration~\citep[e.g.,][]{Murase:2010va,Zhang:2013coa,Yasuda:2024fvc}.    

In this work, we constrain the power of cosmic rays required to explain the observed neutrino fluxes from NGC 1068 considering photohadronic, leptonic and beta decay scenarios.
We explore the parameter space of the injected cosmic-ray or gamma-ray luminosity required to explain the IceCube data in the relevant energy range and revisit constraints on the emission region together with the latest high-energy gamma-ray data from \textit{Fermi}-LAT~\citep{Ajello:2023hkh}.

Our paper is organized as follows. In Section~\ref{sec:model}, we describe the details of the method used in this work. In Sections~\ref{sec:hadronic} and \ref{sec:leptonic}, we present the results related to the photohadronic and leptonic scenarios, respectively. 
In Section~\ref{sec:decay}, we show results related to the beta decay scenario.
We discuss the implications of our results in Section~\ref{sec:dis}.
Finally, we give a summary in Section~\ref{sec:summary}. We use $Q_x = Q/10^x$ in CGS units.

\begin{figure}
    \centering
    \includegraphics[width=\linewidth]{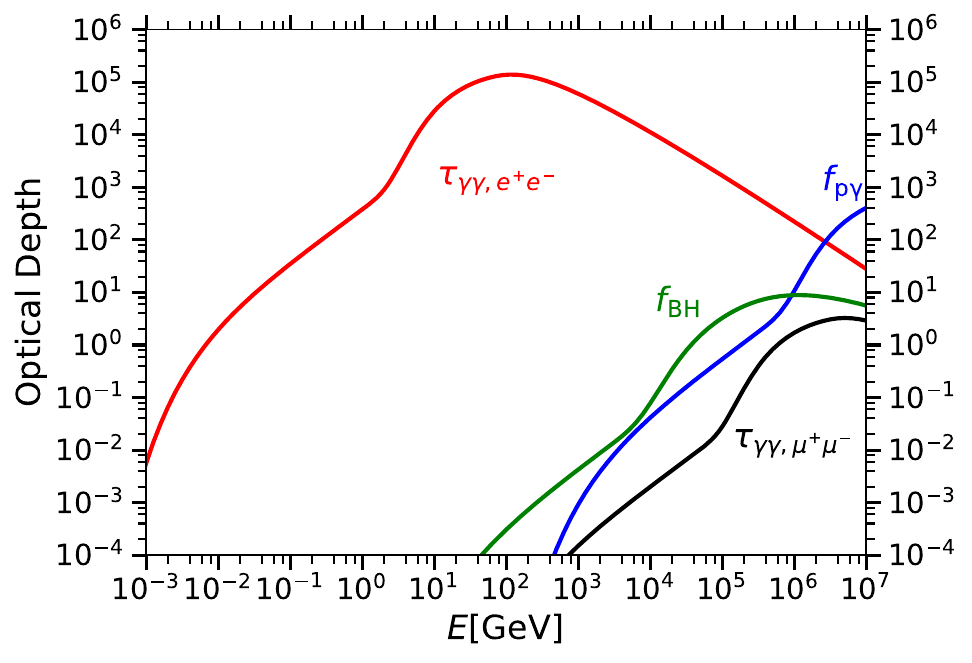}
    \caption{Optical depths for two-photon annihilation, either electron-positron pair production and muon-antimuon pair production, and effective optical depths for photomeson production and Bethe-Heitler pair production. The emission radius is set to $R = 10 R_S$. 
    For the photomeson production and the Bethe-Heitler process, the proton escape timescale is set to $t_* = 10 R/c$.
    }
    \label{fig:optical-depth}
\end{figure}

\section{Method of Calculations}\label{sec:model}
NGC 1068 is an archetypal Seyfert II galaxy that consists of an SMBH with an accretion disk. The luminosity distance is set to $d_L = 10\rm~Mpc$ \citep{Tully:2009ir}. {This value of $d_L$ has been independently inferred by \cite{Courtois:2013yfa} and by \cite{2021AstBu..76..255T}. We use a black hole mass of $M_{\rm BH} = 10^7~\text{M}_\odot$ 
\citep{2002ApJ...579..530W, Panessa:2006sg}, which translates to a Schwarzschild radius $R_S \equiv 2GM_{\rm BH}/c^2 \simeq 3.0 \times 10^{12}$ cm, where $G$ is the gravitational constant and $c$ is the speed of light. 
The Eddington luminosity of NGC 1068 is 
\begin{equation}
L_{\rm Edd}\simeq 1.3\times{10}^{45}~{\rm erg}~{\rm s}^{-1}~\left(\frac{M_{\rm BH}}{10^7 M_\odot}\right),
\end{equation}
and for jet-quiet AGNs the luminosities of cosmic-ray protons, electrons, and nuclei are expected to be lower than this luminosity. 

In this work, we consider high-energy neutrinos and gamma rays produced in the single emission region with size $R$. Based on the previous work, which obtained $R\lesssim30 R_S$~\citep{Murase:2022dog}, for the photohadronic scenario, we restrict the range of $R$ from $0.5R_S$ to $30 R_S$ to evaluate the cosmic-ray luminosity, $L_\text{CR}$, which is required to explain the observed neutrino flux. We consider the similar range of $R$ in the leptonic scenario, because larger radii obviously violate gamma-ray constraints (see below).    
We also assume that the neutrino emission radius is similar to or larger than the X-ray emission region, which is especially reasonable in the corona model for neutrinos. Such a small neutrino emission radius is compatible with the size of magnetically-dominated coronae~\citep[e.g.,][]{Jiang__2019}. In the beta decay scenario, considering the large-scale jet as a cosmic-ray acceleration site \citep{Yasuda:2024fvc}, we assume $R\gtrsim{10}^5~R_S\simeq0.1$~pc outside the dust torus.  

In the photohadronic scenario, neutrinos are produced by the photomeson production, while in the beta decay scenario, neutrinos mainly originate from neutrons produced by the photodisintegration. In both scenarios, the Bethe-Heitler pair production is relevant for electromagnetic emission, especially in the context of AGN coronae~\citep{PhysRevLett.125.011101}. 
The target photon fields within the emission region consist of three components~\citep[e.g.,][]{PhysRevLett.125.011101,Murase:2022dog, Inoue:2022yak}. 
The first component is X-ray emission coming from the hot corona. Considering the distance $d_L = 10$ Mpc, the X-ray luminosity at $2-10$~keV range is $L_{X}^{2-10}=3.38 \times 10^{43}~{\rm erg}~{\rm s}^{-1}$
\citep{Marinucci:2015fqo}. 
The X-ray spectrum is known to be described by a power law with a photon index $\Gamma_{\rm cor} = 2$, which is given by
\begin{equation}
    \varepsilon \frac{dn_{\rm cor}}{d\varepsilon} = \tilde{n}_{\rm cor} {\left(\frac{\varepsilon}{\varepsilon_{\rm cor}^{\rm cut}}\right)}^{1 -\Gamma_\text{cor}}\exp\left(-\frac{\varepsilon}{\varepsilon_{\rm cor}^{\rm cut}}\right),
\end{equation}
where $\tilde{n}_{\rm cor} = \tilde{L}_{\rm cor} / (4\pi R^2 c \varepsilon_{\rm cor}^{\rm cut})$ and $\tilde{L}_{\rm cor} = 2.2\times 10^{43}\rm~erg~s^{-1}$.  
This component is assumed to start at 31.5~eV and has an exponential cutoff energy $\varepsilon_{\rm cor}^{\rm cut}=2kT_{e} = 128$~keV~\citep{Bauer:2014rla}. 
The resulting total coronal luminosity above 2~keV is $L_{\rm cor}^{\rm tot}=7.84\times 10^{43}~\rm~erg~s^{-1}$, and the typical number density of X-ray photons at 2~keV is estimated to be $n_X\approx \tilde{L}_{\rm cor}/(4\pi R^2 c\varepsilon_{X})$, where $\varepsilon_{X}=2$~keV is adopted. 

The second component consists of multitemperature black body emission from the geometrically thin but optically thick disk \citep{Shakura:1972te}.
Using a power-law with a standard index of $\Gamma_{\rm disk} = 2/3$, starting from $\varepsilon_\text{dm} =4.7$~eV and a cutoff energy at $\varepsilon_\text{disk} = 31.5$~eV. This is parameterized as 
\begin{equation}
    \varepsilon \frac{dn_\text{disk}}{d\varepsilon} = \tilde{n}_{\rm disk} \left(\frac{\varepsilon}{\varepsilon_\text{disk}}\right)^{1-\Gamma_{\rm disk}}\exp\left(-\frac{\varepsilon}{\varepsilon_\text{disk}}\right),
\end{equation} 
where $\tilde{n}_{\rm disk} = \tilde{L}_\text{disk} / (4\pi c R^2 \varepsilon_\text{disk})$ and %L_\text{disk,0} = 1.12 L_\text{disk}$, 
%$L_\text{disk} \approx 0.5 \times L_\text{bol} = 5 \times 10^{44}$ erg s$^{-1}$. 
$\tilde{L}_\text{disk} = 5.6 \times 10^{44}$~erg~s$^{-1}$~\citep{Inoue:2022yak}.
Some of the disk emission is reprocessed in the broad-line regions and dust torus, and $L_\text{bol}=4.82\times 10^{44}$~erg~s$^{-1}$ represents the bolometric luminosity~\citep{2002ApJ...579..530W}, implying that the Eddington ratio $\lambda_{\rm Edd}\equiv L_{\rm bol}/L_{\rm Edd}$ should be $\sim0.4-0.5$. 

High-energy gamma rays and electrons/positrons, produced from $p\gamma$ interactions, would further interact with the ambient magnetic field and target photon fields, triggering an electromagnetic cascade process.
We utilize the {\sc Astrophysical Multimessenger Emission Simulator (AMES)} to simulate photohadronic and photonuclear interactions of high-energy protons  and nuclei with target photons within the emission region~\citep[see the details of the method in Appendix A of][]{Zhang:2023ewt}, by which we solve coupled time-dependent kinetic equations for photons, electrons/positrons, neutrinos, protons, neutrons, and nuclei~\citep[see also Appendix of][]{Murase:2017pfe}. 
For cosmic-ray protons and nuclei, we consider the photomeson production, photodisintegration, Bethe-Hitler pair production, and synchrotron energy loss processes. 
We adopt the continuous energy-loss approximation for synchrotron and inverse-Compton processes for electrons/positrons, protons, and nuclei. In the leptonic scenario, {\sc AMES} also allows us to consider the decay products of muon and antimuon pairs, including both neutrinos and electrons/positrons, which are produced from the two-photon annihilation process. 

In this work, the magnetic field energy density is parameterized by the bolometric luminosity, as $U_B = \xi_B U_\gamma$, where $U_B = B^2/8\pi$ is the magnetic energy density and $U_\gamma = L_\text{bol}/(4\pi R^2 c)$ is the bolometric photon energy density. Given that the equipartition parameter is $\xi_B\lesssim1$ for a near-Eddington system such as NGC 1068, we calculate spectra for $\xi_B = $ 1, 0.1, and 0.01.  
We assume no escape for charged particles, while the escape time of neutral particles is set to $t_{\rm esc}=R/c$. We solve the kinetic equations during the dynamical time, $t_{\rm dyn} = R/V$. Although $V = 0.1 c$ is adopted as a default value, we will see that our conclusions are unaffected by this choice. 

Finally, as in \cite{Murase:2022dog}, we take into account gamma-ray attenuation due to matter interactions due to the Bethe-Heitler pair production~\citep{1992ApJ...400..181C} and Compton scatterings, where we use a column density of $N_H=10^{25}~{\rm cm}^{-2}$~\citep{Marinucci:2015fqo}. 
High-energy gamma rays above TeV energies can further be attenuated by infrared emission from the dust torus, whose innermost region has a fiducial radius of $R_\text{DT} = 0.1$~pc~\citep{torus_ALMA1, torus_ALMA2}. The spectrum is approximated by a black body with a temperature of $T_\text{DT} = 1000$~K~\citep{Rosas:2021zbx, Inoue:2022yak} as
\begin{equation}
    \varepsilon\frac{dn_\text{DT}}{d\varepsilon}=\frac{8\pi}{c^3h^3}\frac{\varepsilon^3}{\exp(\varepsilon/k_B T_\text{DT})-1}.
\end{equation}
We multiply cascaded photon spectra by an attenuation factor, $\exp(-\tau_{\gamma\gamma}^{\rm DT})$, where $\tau_{\gamma\gamma}^{\rm DT} = t_{\gamma \gamma}^{-1} R_{\rm DT}/c$ and $t_{\gamma\gamma}$ is the $\gamma \gamma$ interaction timescale in the infrared photon field. 
Note that the energies of the photons making up this component are well below the threshold for neutrino production, and our results on neutrinos are unaffected by the implementation of this component.

\subsection{Optical depths}
The effective optical depth to the Bethe-Heitler pair production process in the disk photon field is estimated to be~\citep[e.g.,][]{PhysRevLett.125.011101}
\begin{eqnarray}\label{eq:BH-loss}
    f_{\rm BH} &\approx& \tilde{n}_{\rm disk} \hat{\sigma}_{\rm BH} R (c / V) \nonumber \\ 
    &\simeq& 7.0 \tilde{L}_{\rm disk, 44.7}
    {\left(\frac{R}{10 R_S}\right)}^{-1}{\left(\frac{M_{\rm BH}}{10^7M_\odot}\right)}^{-1} \nonumber \\ 
    &\times& {\left(\frac{\varepsilon_{\rm disk}}{31.5\rm~eV}\right)}^{-1} {\left(\frac{V}{0.1 c}\right)}^{-1},
\end{eqnarray}
where $\hat{\sigma}_{\rm BH} \sim 0.8 \times 10^{-30}\rm~cm^2$ and the typical proton energy causing pair production is $\tilde{\varepsilon}_{p}^{\rm BH-disk} \approx 0.5 \bar{\varepsilon}_{\rm BH} m_p c^2 / \varepsilon_{\rm disk} \simeq 1.5 \times 10^5\rm~GeV~(\varepsilon_{\rm disk} / 31.5\rm~eV)^{-1}$ and $\bar{\varepsilon}_{\rm BH} \approx 10\rm~MeV$.
In the photohadronic scenario, TeV neutrinos mostly come from interactions with X-ray photons, and the effective optical depth to the photomeson production process is estimated to be
\begin{eqnarray}\label{eq:pg-loss}
    f_{p\gamma} &\approx& \eta_{p\gamma} \hat{\sigma}_{p\gamma} R (c /V) \tilde{n}_X \left(\frac{\varepsilon_p}{\varepsilon_{p}^{p\gamma-X}}\right)^{\Gamma_{\rm cor} - 1} \nonumber \\ 
    &\simeq& 0.39~\eta_{p\gamma} \tilde{L}_{\rm cor, 43.3}\left(\frac{R}{10 R_S}\right)^{-1}{\left(\frac{M_{\rm BH}}{10^7M_\odot}\right)}^{-1}\nonumber\\
    &\times&\left(\frac{V}{0.1 c}\right)^{-1} {\left(\frac{\varepsilon_p}{\varepsilon_p^{p\gamma-X}}\right)}^{\Gamma_{\rm cor} - 1},
\end{eqnarray}
where $\eta_{p\gamma} = 2 / (1 + \Gamma_{\rm cor})$, $\hat{\sigma}_{p\gamma} \sim 0.7\times 10^{-28}\rm~cm^2$ and $\varepsilon_{p}^{p\gamma-X} \simeq 7.9 \times 10^4\rm~GeV~(\varepsilon_{X}/2\rm~keV)^{-1}$.
We can see that, at $\sim 10\rm~TeV$, proton energy losses via the Bethe-Heitler pair production process are more significant than the photomeson production, which is consistent with Fig.~\ref{fig:optical-depth}. 

The optical depths to electron-positron and muon-antimuon pair production from the two-photon annihilation process are shown in Fig.~\ref{fig:optical-depth}.
For the electron-positron pair production, the threshold gamma-ray energy is $\tilde{\varepsilon}_\gamma^{e^+e^--X} \approx m_e^2 c^4 / \varepsilon_X \simeq 0.13\rm~GeV~(\varepsilon_X/2\rm~keV)^{-1}$, and the two-photon annihilation optical depth to electron-positron pair production in the GeV gamma-ray range is~\citep{Murase:2022dog}
\begin{align}
    \tau_{\gamma \gamma \to e^+ e^-} &\approx 
    \eta_{\gamma \gamma} \sigma_T R \tilde{n}_{X} \left(\frac{\varepsilon_\gamma}{\tilde{\varepsilon}_\gamma^{e^+e^--X}}\right)^{\Gamma_{\rm cor} - 1} \nonumber \\ &\simeq 44~\tilde{L}_{\rm cor, 43.3}\left(\frac{R}{10 R_S}\right)^{-1} {\left(\frac{M_{\rm BH}}{10^7M_\odot}\right)}^{-1} \nonumber \\ &\quad \left(\frac{\varepsilon_\gamma}{\tilde{\varepsilon}_\gamma^{e^+e^--X}}\right)^{\Gamma_{\rm cor} - 1},
\end{align}
where $\eta_{\gamma\gamma}\approx 0.12$ is the coefficient~\citep{sve87} and $\sigma_T \approx 6.65\times 10^{-25}\rm~cm^{2}$ is the Thomson cross section.
The two-photon annihilation optical depth to muon-antimuon pair production is
\begin{eqnarray}
    \tau_{\gamma \gamma \to \mu^+ \mu^-} &\approx& 
    \eta_{\gamma \gamma} (m_e/m_\mu)^2 \sigma_T R \tilde{n}_{X} \left(\frac{\varepsilon_\gamma}{\tilde{\varepsilon}_\gamma^{\mu^+\mu^--X}}\right)^{\Gamma_{\rm cor} - 1} \nonumber \\ 
    &\simeq& 1.0 \times 10^{-3} ~\tilde{L}_{\rm cor, 43.3} \left(\frac{R}{10 R_S}\right)^{-1} {\left(\frac{M_{\rm BH}}{10^7M_\odot}\right)}^{-1} 
    \nonumber \\ 
    &\times& \left(\frac{\varepsilon_\gamma}{\tilde{\varepsilon}_\gamma^{\mu^+\mu^--X}}\right)^{\Gamma_{\rm cor} - 1},
\end{eqnarray}
where $\tilde{\varepsilon}_\gamma^{\mu^+\mu^--X} \approx m_\mu^2 c^4 / \varepsilon_{X} \simeq 5.6\times 10^3\rm~GeV~(\varepsilon_{X}/2~{\rm keV})^{-1}$. 
The above expressions for $\tau_{\gamma \gamma \to e^+ e^-}$ and $\tau_{\gamma \gamma \to \mu^+ \mu^-}$ are consistent with the numerical results shown in Fig.~\ref{fig:optical-depth}. We can also see the optical depth to the electron-positron pair production at $\sim10$~TeV is $\sim 10^6$ larger than that to the muon-antimuon pair production.

\section{Photohadronic scenario}\label{sec:hadronic}
We assume that the injected cosmic-ray proton spectrum follows a power-law distribution with an exponential cutoff,
\begin{equation}
\frac{d\dot{n}_\text{CR}}{d\varepsilon_p} \propto \varepsilon_p^{-s_{\rm CR}} {\rm exp}\left({-\frac{\varepsilon_p}{\varepsilon_p^\text{max}}}\right),
\end{equation}
where $s_{\rm CR}$ is the proton power-law index and $\varepsilon_p^\text{max}$ is the proton maximum energy. 
The normalization is set by the cosmic ray proton luminosity which is
\begin{equation}
L_\text{CR} = \mathcal{V} \int_{\varepsilon_p^\text{min}}^{\infty} \varepsilon_p \frac{d\dot{n}_\text{CR}}{d\varepsilon_p} d\varepsilon_p,
\end{equation} 
where $\mathcal{V} = (4\pi/3) R^3$ is the volume of the emission region.
%from the 10-year IceCube data which is equal to $(2.9 \pm 1.1) \times 10^{42}~\text{erg~ s}^{-1}$~\citep{IceCube:2022der}. 
We perform simulations for $\varepsilon_p^\text{min}=10$~TeV and $\varepsilon_p^\text{max}=30$~TeV, but vary the spectral index, $s_{\rm CR}$, from 1 to 4.
Below $\varepsilon_p^{\rm min} = 10\rm~TeV$, a sufficiently hard spectrum is required not to violate the energetics requirement~\citep{Murase:2022dog}. Since the observed neutrino flux is extended down to $\varepsilon_\nu \sim 1 \rm~TeV$, corresponding to a parent proton energy of $\varepsilon_p \sim 20 \varepsilon_\nu \sim 20\rm~TeV$, it is reasonable to adopt a low-energy cutoff at $\varepsilon_p^{\rm min} = 10\rm~TeV$ to evaluate the minimum cosmic-ray proton luminosity. The value of $\varepsilon_p^{\rm max}=30\rm~TeV$ is a conservative choice as the high-energy cutoff for the purpose of evaluating $L_{\mr CR}$. A larger value results in more neutrinos with higher energies, which in turn requires a larger value of $L_\text{CR}$, while the $1.5-15$~TeV neutrino flux may match the IceCube data.

The injected cosmic-ray luminosity, $L_{\rm CR}$, is constrained by comparing the all-flavor neutrino luminosity between 1.5~TeV and 15~TeV to the value obtained by the IceCube collaboration, $L_{\nu} = (1.4 \pm 0.53) \times 10^{42}\rm~erg~s^{-1}$~\citep{IceCube:2022der}.
Note that we set the luminosity distance to $d_L = 10$~Mpc, which is shorter than the luminosity distance $d_L = 14.4\rm~Mpc$ used in ~\cite{IceCube:2022der}.

\subsection{Multimessenger spectra and gamma-ray constraints}
\begin{figure*}
    \centering
    \includegraphics[width=\linewidth]{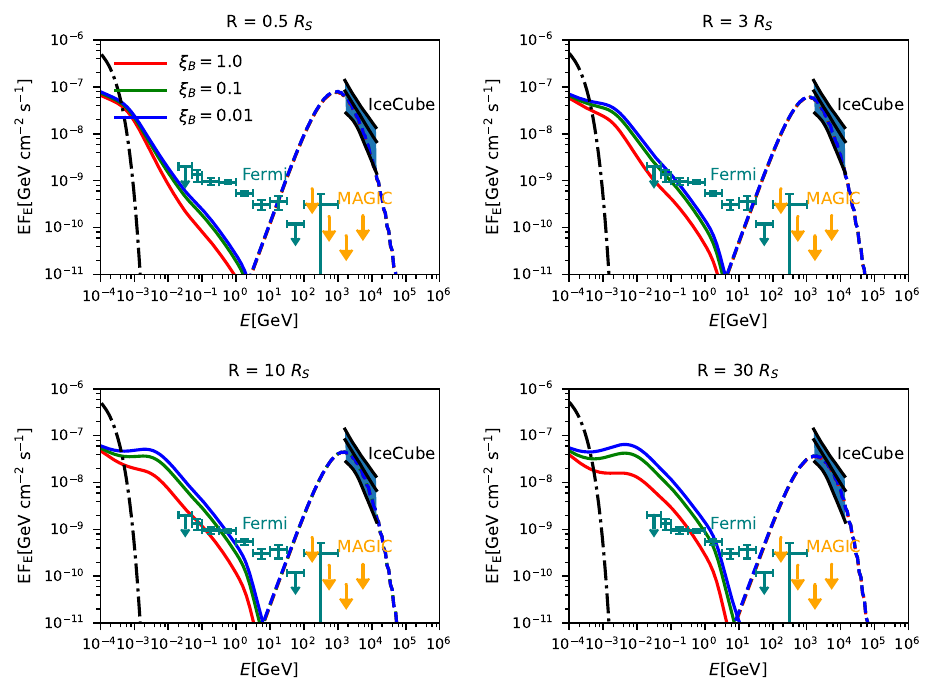}
    \caption{Cascaded photon spectra (solid curves) and all-flavor neutrino spectra (dashed lines curves) for different values of $R$ and $\xi_B$ in the photohadronic scenario. All plots are made using parameters for injected cosmic-ray protons, $s_{\rm CR}=2$, $\varepsilon_{p}^{\rm min}=10$~TeV, and $\varepsilon_{p}^{\rm max}=30$~TeV. The black lines correspond to the 95 percent contour lines and line of best fit from the IceCube data~\citep{IceCube:2022der}. Gamma-ray data from the \emph{Fermi}-LAT~\citep{Ajello:2023hkh} and MAGIC~\citep{MAGIC:2019fvw} observations are also shown. The tail of the intrinsic X-ray flux used in the calculation is represented by the dash-dotted curves.}
    \label{fig:spectrum-plots}
\end{figure*}

We show cascaded photon and all-flavor neutrino spectra in Fig.~\ref{fig:spectrum-plots} for different emission radii with $R = 0.5, 3, 10, 30R_S$, respectively.
The neutrino spectra are shown with dashed lines, while the cascade photon spectra are shown with solid lines. We also depict the intrinsic flux of X-rays with dotted lines. Note that the cascade flux in the X-ray range is lower than the intrinsic X-ray flux from coronae, as shown in previous works~\citep{PhysRevLett.125.011101,Murase:2022dog}, so that constraints from X-ray observations are much weaker than those from gamma-ray observations. The AGN corona model itself predicts that the neutrino luminosity is approximately proportional to the X-ray luminosity~\citep{Murase:2016gly,PhysRevLett.125.011101} and the cosmic-ray luminosity is expected to be lower than the X-ray luminosity~\citep{Murase:2015xka}.

The cascaded photon spectra vary with emission radii. For larger emission radii, e.g., $R = 30 R_S$, the cascade photon spectra overshoot the \textit{Fermi}-LAT data obtained from sub-GeV to GeV energies~\citep{Ajello:2023hkh}, especially for $\xi_B = 0.1$ and $\xi_B = 0.01$. While our numerical results confirm those of \cite{Murase:2022dog}, the gamma-ray data adopted in this work~\citep{Ajello:2023hkh} extend to lower and higher energies, which give tighter constraints on $R$. The synchrotron cascade with $\xi_B \gtrsim 0.1$ allows more parameter space for $R$, because more gamma rays appear at sub-GeV or lower energies.

Then, we obtain gamma-ray constraints on emission radii by requiring the cascade photon spectra from our simulations not to overshoot the measured gamma-ray data, especially the lowest-energy {\it Fermi}-LAT upper limit at the $95\%$ confidence level. 
Our results are represented by the red and orange curves in Fig.~\ref{fig:contour-plot-hadronic}, which are calculated for $\xi_B = 1.0$ and $\xi_B = 0.01$, respectively. 
We find that the size of the emission region is constrained to $R \lesssim 15 R_S$ for $\xi_B = 1.0$ and to $R \lesssim 3 R_S$ for $\xi_B = 0.01$. The constraints are tighter for softer spectral indices of the injected proton spectrum. This is because a softer cosmic-ray spectrum results in more Bethe-Heitler pair production processes that give rise to larger cascaded photon fluxes.

Our constraints on $R$ for $\xi_B\lesssim0.01$ imply that the neutrino emission region is unlikely to be high-$\beta$ plasma if NGC 1068 is a Schwarzschild black hole. This is because the upper limits on $R$ are smaller than the innermost stable circular orbit (ISCO) for the non-rotating case. While it does not exclude ISCOs for a Kerr black hole with prograde orbits, given that NGC 1068 does not have a strong jet, an extreme Kerr black hole might also be disfavored although details depend on the magnetic flux configuration.

\subsection{Required power from neutrino observations}
\begin{figure}
    \centering  
    \includegraphics[width=0.5\textwidth]{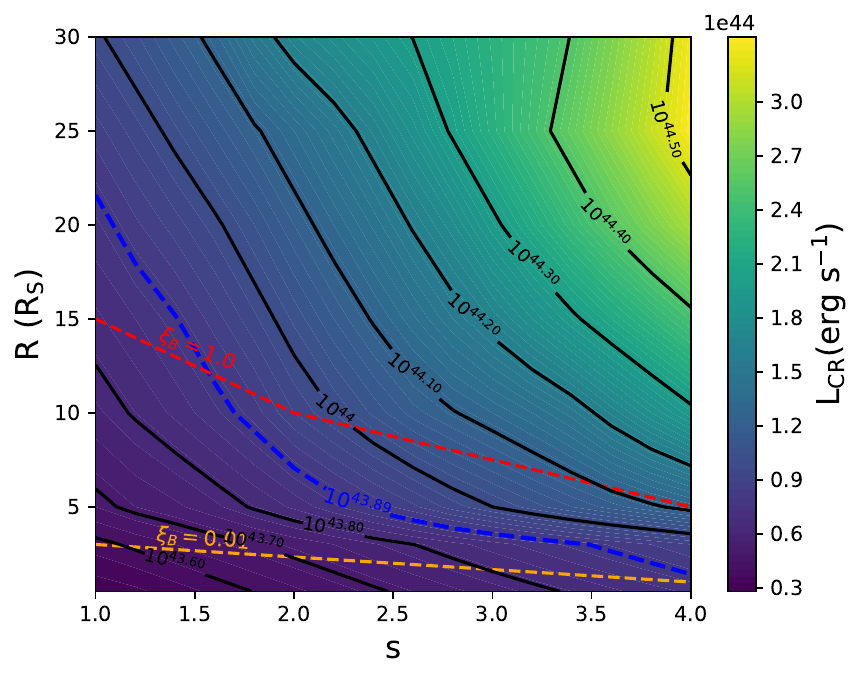}
    \caption{Required minimum cosmic-ray proton luminosity, $L_\text{CR}$, as a function of the emission radius $R$ and power-law index $s_{\rm CR}$ for $\varepsilon_p^{\rm min}=10$~TeV, $\varepsilon_p^{\rm max}=30$~TeV, and $\xi_B$ = 1.0 in the photohadronic scenario. Regions above the red and orange dashed lines are excluded by the \emph{Fermi}-LAT data~\citep{Ajello:2023hkh} for $\xi_B = 1.0$ and $\xi_B = 0.01$, respectively. The blue contour corresponds to $L_{\rm cor}^{\rm tot}$, which is the total coronal luminosity.}
    \label{fig:contour-plot-hadronic}
\end{figure}

In Fig.~\ref{fig:contour-plot-hadronic}, we show the required cosmic-ray proton luminosity as a function of $R$ and $s_{\rm CR}$.
We find that the typical range of the injected cosmic-ray luminosity for our chosen parameter space is $L_\text{CR} \sim (3 - 30)
\times 10^{43}$~erg~s$^{-1}$, which is less than the bolometric and Eddington luminosities of NGC 1068. 
From Eq.~\ref{eq:pg-loss}, we see the efficiency for neutrino production is proportional to $R^{-1}$, which means smaller radii are preferred for efficient neutrino production. 
Correspondingly, the required minimum cosmic-ray luminosity may decrease with emission radii. 

On the other hand, harder spectral indices could alleviate the required cosmic-ray power. While the IceCube data provide the best-fit spectrum with a spectral index of 3.2~\citep{IceCube:2022der}, a hard spectra with a cutoff may explain the data \citep{Murase:2022dog}. For $s_\text{CR}\lesssim 2$, protons with energy around $\varepsilon_p^\text{max}$ are dominant. Therefore, the $L_\text{CR}$ requirement changes drastically upon going from $s_\text{CR}=1$ to $s_\text{CR}=2$ as it mostly increases the density of lower-energy protons that do not contribute towards neutrino production. 
For $s_\text{CR}\lesssim 2$, protons with energy around $\varepsilon_p^\text{min}$ are dominant. Therefore, further increasing the softness does not affect the required $L_\text{CR}$ much. 

The minimum cosmic-ray proton power is estimated to be
\begin{equation}
L_{\rm CR} \gtrsim (3-30)\times{10}^{43}~{\rm erg}~{\rm s}^{-1},  
\end{equation}
which can be below the bolometric and Eddington luminosities, so we conclude that the photohadronic scenario is viable as a mechanism for neutrino emission from NGC 1068.

\section{Leptonic scenario}\label{sec:leptonic}
In principle, neutrinos could be produced by a pure leptonic process through the two-photon annihilation and muon-antimuon pair production~\citep[e.g.,][as an example for neutrino production in intergalactic space]{li2007eev}. This process can be relevant only at sufficiently high energies, where electron-positron pair production is highly suppressed by Klein-Nishina-like effects. The decay of muons and antimuons will then produce neutrinos without involving the acceleration of cosmic-ray protons and nuclei. Assuming that very high-energy gamma-rays are produced, \cite{Hooper:2023ssc} argued that primary electrons with a luminosity of $L_e \sim 10^{42}-10^{43}$~erg~s$^{-1}$ injected above 1~TeV is needed to explain the observed neutrino flux. In addition, considering kG-scale magnetic fields and dense keV-scale target photons, they adopted a black body spectrum with a temperature of $T = 1 - 10$~keV as a target photon field.  

In this work, we consider the realistic target photon fields discussed in Sec.~\ref{sec:model} to test this scenario. We will show that the multiwavelength data observed at the low-energy gamma-ray band \citep{Ajello:2023hkh, MAGIC:2019fvw} give strong constraints on this scenario, considering that the associated electromagnetic cascade process is unavoidable. 
For the purpose of this work, we inject a power-law spectrum of high-energy gamma rays with an exponential cutoff at both low- and high-energy parts, which is written as
\begin{equation}
    \frac{d\dot{n}_\gamma}{d\varepsilon_\gamma} \propto  
    \varepsilon_\gamma^{-s_\gamma} {\rm exp}\left({-\frac{\varepsilon_\gamma^\text{min}}{\varepsilon_\gamma}}\right){\rm exp}\left({-\frac{\varepsilon_\gamma}{\varepsilon_\gamma^\text{max}}}\right),
\end{equation}
where $\varepsilon_\gamma^\text{min}$ = 12~TeV and $\varepsilon_\gamma^\text{max}$ = 15~TeV.
As in the photohadronic scenario, we consider a range of spectral index of the injected gamma-ray spectrum, $s_\gamma$, between 1 and 4. The total injected gamma-ray luminosity is given by
\begin{equation}
L_\gamma = \mathcal{V} \int_{-\infty}^{\infty}  d\varepsilon_\gamma  \frac{d\dot{n}_\gamma}{d\varepsilon_\gamma} \varepsilon_\gamma.
\end{equation}

\subsection{Multimessenger spectra and gamma-ray constraints}
\begin{figure*}
    \centering
    \includegraphics[width=\linewidth]{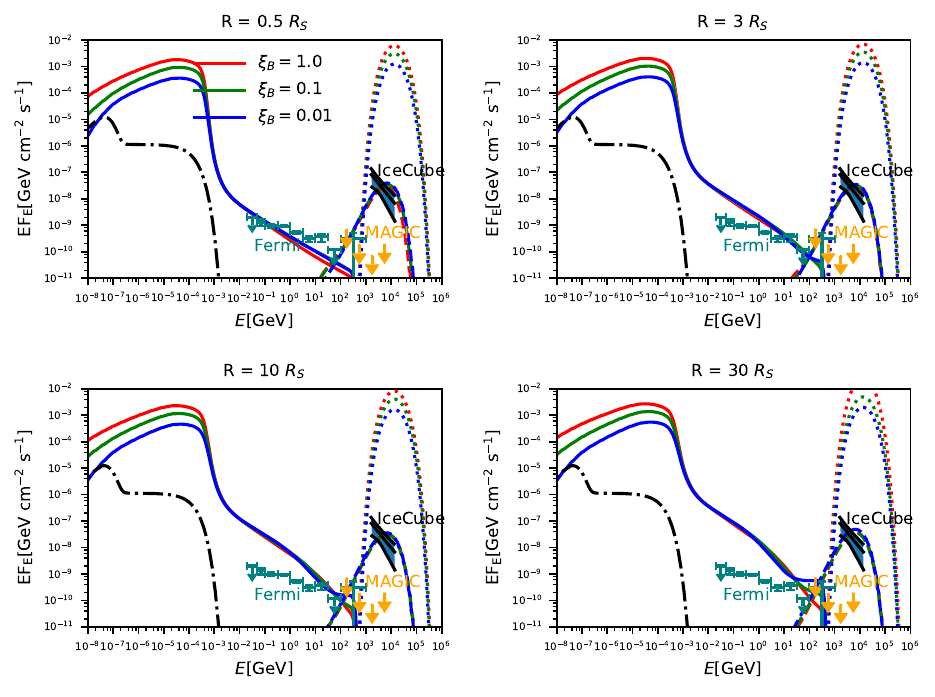}
    \caption{Same as in Fig.~\ref{fig:spectrum-plots}, but we show cascaded photon spectra and all-flavor neutrino spectra (solid and dashed lines respectively) for different values of $R$ and $\xi_B$ in the leptonic scenario. All panels are made using parameter values $s_\gamma$ = 2.0, $\varepsilon_{\gamma}^{\rm min}= 12$~TeV and $\varepsilon_{\gamma}^{\rm max}=15$~TeV for the injected photon spectrum.}
    \label{fig:result-lepto}
\end{figure*}

In Fig.~\ref{fig:result-lepto}, we show cascaded photon and all-flavor neutrino spectra in the leptonic scenario (solid and dashed lines, respectively). The injected gamma-ray spectra are shown using dotted lines. The injected gamma rays interact with the disk and coronal photons, and some of them produce muon-antimuon pairs.
Most of gamma rays are cascaded down to MeV or lower energies, and overwhelm the intrinsic coronal spectrum. 

We also examine cascade constraints as in the photohadronc scenario. 
In the leptonic scenario, we find that the cascaded photon spectra violate gamma-ray observations including the \emph{Fermi}-LAT data~\citep{Ajello:2023hkh} for almost all the region of parameter space that we test. Also, the cascaded X-ray flux is much larger than the intrinsic X-ray flux inferred by {\it NuSTAR} observations (represented by the dotted-dash line). We do not find viable parameter space for the leptonic scenario to explain the observed IceCube data~\citep{IceCube:2022der} without violating the intrinsic X-ray or gamma-ray data.

This situation can be understood as follows. At $\varepsilon_\gamma \gtrsim {\rm a~few}~\rm~GeV$, the two-photon annihilation optical depth for electron-positron pair production is expected to be dominated by the interaction with disk photons or other optical and ultraviolet photons if they exist. The optical depth at $\gtrsim10-100$~GeV is naively estimated by the following expression, 
\begin{align}
    \tau_{\gamma \gamma \to e^+ e^-} 
    %&\approx 
    %\eta_{\gamma \gamma} \sigma_T R n_{\rm disk} \nonumber \\ 
    &\simeq 1.1 \times 10^6~
    {\left(\frac{\varepsilon_\gamma^{e^+e^--{\rm dm}}}{\tilde{\varepsilon}_\gamma^{e^+e^--{\rm disk}}}\right)}^{-1/3}
    \left(\frac{\varepsilon_\gamma}{\tilde{\varepsilon}_\gamma^{e^+e^--{\rm dm}}}\right)^{-1}\nonumber \\ &\times
    \tilde{L}_{\rm disk, 44.7} \left(\frac{R}{10 R_S}\right)^{-1}{\left(\frac{M_{\rm BH}}{10^7M_\odot}\right)}^{-1} \Lambda_1,
  %{\rm ln}\left( \frac{0.4\varepsilon_\gamma}{\tilde{\varepsilon}_{\gamma\gamma, e^+e^-}}\right),
\end{align}
where 
% $x = \varepsilon_\gamma / \tilde{\varepsilon}_{\gamma\gamma, e^+e^-}$ and
$\tilde{\varepsilon}_\gamma^{e^+e^--{\rm disk}} \approx m_e^2 c^4 / \varepsilon_{\rm disk} \simeq 8.3\rm~GeV~(\varepsilon_{\rm disk} / 31.5\rm~eV)^{-1}$, $\tilde{\varepsilon}_\gamma^{e^+e^--{\rm dm}} \approx m_e^2 c^4 / \varepsilon_{\rm dm} \simeq 56\rm~GeV~(\varepsilon_{\rm dm} /4.7\rm~eV)^{-1}$, and $\Lambda$ is the logarithmic factor. In the leptonic scenario, neutrinos in the TeV range are mostly produced via interactions with coronal photons or other high-energy photons originating from electromagnetic cascades. 

For given $\varepsilon_X$ and $\varepsilon_{\rm disk}$, with $\Gamma_{\rm cor}=2$, the ratio of the above two optical depths at sufficiently high energies is estimated by
\begin{eqnarray}
\frac{\tau_{\gamma \gamma \to e^+ e^-}}{\tau_{\gamma \gamma \to \mu^+ \mu^-}} &\sim2\times10^6~\left(\frac{\tilde{L}_{\rm disk, 44.7}}{\tilde{L}_{\rm cor, 43.3}}\right) \Lambda_1,   
\end{eqnarray}
which roughly agrees with the numerical result shown in Fig.~\ref{fig:optical-depth}. This ratio is also consistent with that of the gamma-ray flux to the neutrino flux shown in Fig.~\ref{fig:result-lepto}. For comparison, \cite{Hooper:2023ssc} argued that the ratio is $\sim 0.1 - 0.01$ for the optical depth at around 10~TeV assuming a black body spectrum with a temperature of $1\rm~keV$. For a realistic X-ray spectrum that is described by a power law, as shown in this work, the ratio is much larger, which is the reason why the contribution of neutrinos from muon-antimuon production was not usually shown in the previous calculations~\citep{Murase:2022dog}.

\subsection{Required power from neutrino observations}
\begin{figure}
    \centering    
    \includegraphics[width=0.5\textwidth]{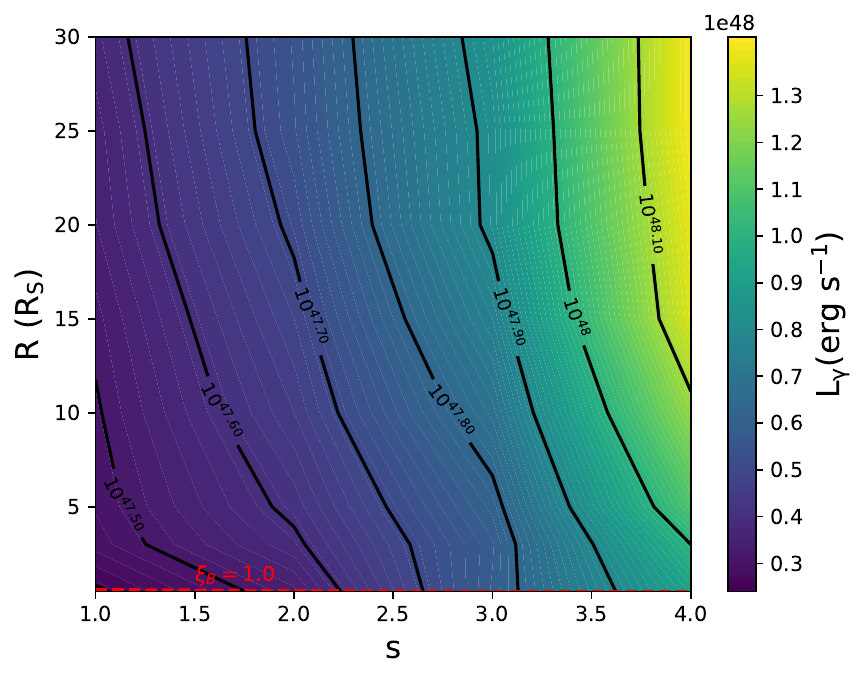}
    \caption{Required minimum gamma-ray luminosity, $L_\gamma$, as a function of $R$ and $s_\gamma$, for $\varepsilon_\gamma^{\rm min}=12$~TeV, $\varepsilon_\gamma^{\rm max}= 15$~TeV, and $\xi_B$ = 1.0, in the leptonic scenario. We find that $L_{\rm CR}$ exceeds $L_{\rm bol}\sim L_{\rm Edd}$ in all the parameter space. The region above the red dashed line is excluded by the \emph{Fermi}-LAT data~\citep{Ajello:2023hkh} for $\xi_B = 1.0$. All parameter space is excluded for $\xi_B = 0.01$.}
    \label{fig:contour-lepto}
\end{figure}

As seen in Fig.~\ref{fig:contour-lepto}, we calculate the required power of injected gamma rays, which is $L_\gamma\sim (1-4)\times 10^{47}$~erg~s$^{-1}$ to explain the observed neutrino flux in the leptonic scenario,  
These values are about two orders of magnitude greater than the bolometric luminosity. Our results on the required power of injected gamma rays are also about four orders larger than the required electron luminosity derived in \cite{Hooper:2023ssc}. The difference is that we adopt a realistic coronal photon field with a differential energy density of $\tilde{U}_{\rm cor}\simeq5.9\times 10^4~{\rm~erg~cm^{-3}}~(R/10 R_S)^{-2}\tilde{L}_{\rm cor,43.3}$, which is many orders smaller than the energy density of the black body photon field with temperate $T = 1\rm~keV$. 

Compared to the required cosmic-ray power in the photohadronic scenario, the required power of injected gamma rays in the leptonic scenario is insensitive to the size of the emission region, as shown in Fig.~\ref{fig:contour-plot-hadronic}. 
The reason is that, for environments that are optically thick to electron-positron pair production, the neutrino flux is determined by the ratio between muon-antimuon and electron-positron pair production. Analytically, considering $\tau_{\gamma \gamma \to e^+ e^-}\gg1$, the all-flavor neutrino flux is estimated by
\begin{eqnarray}
\varepsilon_\nu \frac{dL_{\varepsilon_\nu}}{d\varepsilon_\nu}
\approx 
\frac{2}{3}\frac{\tau_{\gamma \gamma \to \mu^+ \mu^-}}{\tau_{\gamma \gamma \to e^+ e^-}}\varepsilon_\gamma \frac{dL_{\varepsilon_\gamma}}{d\varepsilon_\gamma}.
\end{eqnarray}

Given that the ratio is $\sim{10}^5-{10}^6$, the minimum gamma-ray power is 
\begin{equation}
L_\gamma \sim ({10}^5-{10}^6)L_\nu \sim ({10}^{47}-{10}^{48})~{\rm erg}~{\rm s}^{-1} \gg L_{\rm bol},  
\end{equation}
The required minimum gamma-ray luminosity is much larger than the Eddington luminosity, so that we conclude that the muon-antimuon pair production process is excluded as a mechanism for high-energy neutrinos from NGC 1068.

\section{Beta decay scenario}\label{sec:decay}
We assume that the injected cosmic-ray nuclei spectrum follows a power-law distribution with an exponential cutoff,
\begin{equation}
\frac{d\dot{n}_\text{CR}}{d\varepsilon_A} \propto \varepsilon_A^{-s_{\rm CR}} {\rm exp}\left({-\frac{\varepsilon_A}{\varepsilon_A^\text{max}}}\right),
\end{equation}
where $s_{\rm CR}$ is the power-law index and $\varepsilon_A^\text{max}$ is the maximum energy of cosmic-ray nuclei. For the composition, we assume helium nuclei but the results are similar for heavier nuclei. 
The normalization is set by the cosmic-ray luminosity which is
\begin{equation}
 L_\text{CR} = \mathcal{V} \int_{\varepsilon_A^\text{min}}^{\infty} \varepsilon_A \frac{d\dot{n}_\text{CR}}{d\varepsilon_A} d\varepsilon_A.
\end{equation} 
We adopt the same target photon fields as in the photohadronic and leptonic scenarios. 
In the beta decay scenario, as in \cite{Yasuda:2024fvc}, we assume that the emission region is located outside the dust torus, which is justified if nuclei are accelerated in jets \citep{Pe'er:2009rc}. We consider not only infrared photons from the dust torus but also the cosmic microwave background (CMB) and extragalactic background light (EBL). 
For target photons from the dust torus, the energy density is assumed to decrease as $(R_{\rm DT}/R)^{2}$ at $R > R_{\rm DT}$. 
The mean lifetime of free neutrons is $\tau_n \approx 879.6 \rm~s$~\citep{ParticleDataGroup:2022pth}, and we consider beta decay neutrinos from both inside and outside the emission region.
%To account for the decay process of escaping neutrons outside the emission region, we artificially adopt a much smaller neutron rest frame decay time. 
The magnetic field strength in the emission region is assumed to be $B = 1~\mu \rm G$ as a default value, but we checked that our conclusions are largely insensitive to this choice because the energy density of the dust component is dominant. For the EBL, we use the model of \cite{Gilmore:2011ks}, and take into account gamma-ray attenuation due to CMB and EBL during intergalactic propagation.

\subsection{Multimessenger spectra and gamma-ray constraints}
In Fig.~\ref{fig:result-decay}, we show cascaded photon and all-flavor neutrino spectra in the beta decay scenario (solid and dashed lines, respectively). The injected cosmic-ray spectra are shown with dotted lines. 
The blue lines consider the photodisintegration process, the photomeson production process, the Bethe-Heitler pair production process, and neutron decay.
For comparison purposes, we also show the case without considering the photomeson production process with red lines.

The effective photodisintegration cross section for $A\leq4$ (including inelasticity with $1/A$) is given by $\hat{\sigma}_{\rm dis}\approx 4.3\times{10}^{-28}~{\rm cm}^2~{(A/4)}^{-2.433}$~\citep{Karakula:1994nv,Murase:2010gj}. Then, the effective optical depth to photodisintegration for helium nuclei is 
\begin{eqnarray}
f_{\rm dis}&\approx&\tilde{n}_{\rm disk}\hat{\sigma}_{\rm dis}R (c/V)
\nonumber \\
&\simeq& 3.8\times{10}^{-4}~\tilde{L}_{\rm disk, 44.7} \left(\frac{R}{10^8 R_S}\right)^{-1}{\left(\frac{M_{\rm BH}}{10^7M_\odot}\right)}^{-1}\nonumber\\
&\times& \left(\frac{\varepsilon_{\rm disk}}{31.5\rm~eV}\right)^{-1}  \left(\frac{V}{0.1 c}\right)^{-1},
\end{eqnarray}
which can be almost unity at sufficiently small emission radii. Note that a fraction of energy carried by nucleons is only $1/A$ per collision, which should be considered for the calculations of neutrinos and gamma rays~\citep[e.g.,][]{Murase:2010va,Zhang:2023ewt}. 
In addition, the average energy of neutrinos from neutron decay is $0.48$~MeV in the neutron rest frame~\citep[e.g.,][]{Murase:2010va}, so the energy fraction carried by neutrinos is $0.48~{\rm MeV}/(m_nc^2)\sim1/(2000)$. The average kinetic energy of beta decay electrons is $0.3$~MeV~\citep{Zhang:2023ewt}. 

One important point is that the Bethe-Heitler pair production process is unavoidable. This is because the threshold for electron-positron production, $2m_ec^2$, is lower than that for nuclear disintegration. The similar efficiency issue exists for deexcitation gamma rays although the energy range of deexcitation gamma rays is different from cascade gamma rays from the Bethe-Heitler pairs and beta decay electrons~\citep{Murase:2010va,Aharonian:2010te,Zhang:2023ewt}.
When disk photons are main targets, the ratio of the Bethe-Heitler contribution to the beta decay contribution is given by
\begin{equation}
\frac{6000f_{\rm BH}(Z^2/A)}{f_{\rm dis}}\sim 10,  
\end{equation}
where $Z^2/A=1$ for helium and $f_{\rm BH}$ in Eq.~\ref{eq:BH-loss} is defined for protons. 
While we find that the cascaded photon spectrum only from beta decay electrons overshoots the gamma-ray data including the {\it Fermi}-LAT data and MAGIC upper limits, this conclusion is robust in the presence of electron-positron pairs from the Bethe-Heitler process. 

Note that the electromagnetic cascade is likely to be dominated by inverse-Compton emission. The dust field is dominant over CMB and EBL for sufficiently small emission radii that assure efficient photodisintegration, at which the magnetic field is weaker than the equipartition magnetic field, 
\begin{eqnarray}
B_{\rm eq}&=&\sqrt{\frac{8\pi L_{\rm DT}}{4\pi R^2 c}}\nonumber\\
&\simeq&0.24~{\rm G}~{(R/R_{\rm DT})}^{-1}{(0.1~{\rm pc}/R_{\rm DT})}L_{\rm DT,43.9}^{1/2}.
\end{eqnarray}
We have also checked the results with a higher value of the magnetic field strength with $B = 300\rm~\mu G$, and our conclusions remain unchanged. Even though the results remain similar, the cascade photon flux may also be enhanced because of the contribution from synchrotron emission. 
%when increasing the emission radius.

Furthermore, the primary gamma-ray component from $\pi^0$ decay must exist once the photomeson production is consistently included. The threshold energy for the pion production is higher, but due to the higher efficiency~\citep[see also][]{Murase:2010va,Murase:2010gj} we find that the contribution from the photomeson production is not negligible and violates the MAGIC upper limits independent of magnetic fields. 

\begin{figure*}
    \centering
    \includegraphics[width=\linewidth]{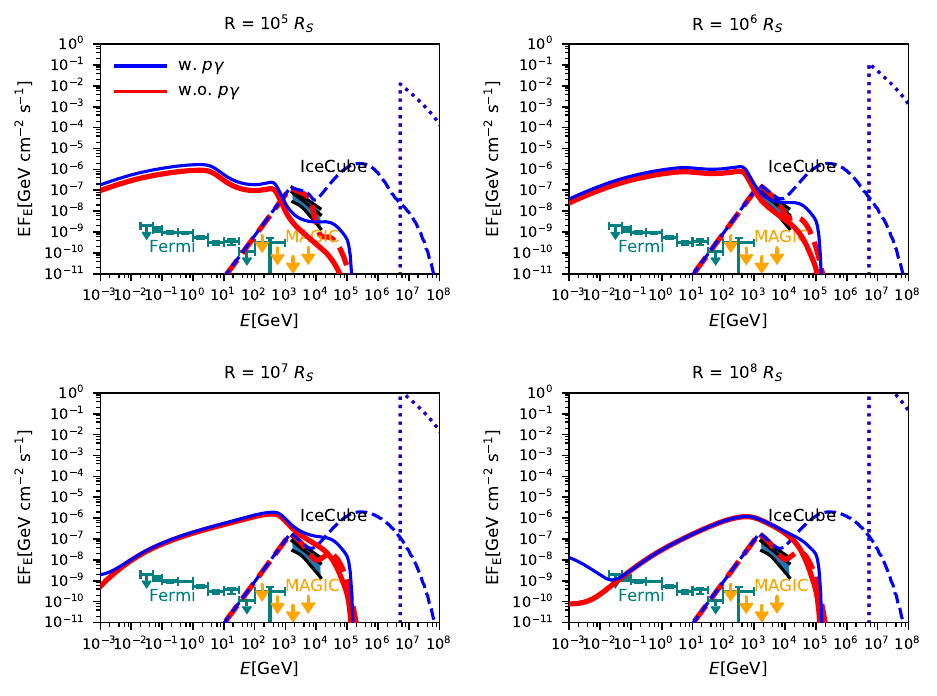}
    \caption{Same as in Fig.~\ref{fig:spectrum-plots}, but we show the cascaded gamma-ray spectra and all-flavor neutrino spectra (solid and dashed lines respectively) for different values of $R$ in the decay scenario. The EBL attenuation is considered~\citep{Gilmore:2011ks}. All panels are made using parameter values $s_{\rm CR}$ = 3.2, $\varepsilon_{A}^{\rm min}$ =5~PeV and $\varepsilon_{A}^{\rm max}$ = 100~PeV for the injected helium nuclei spectrum.}
    \label{fig:result-decay}
\end{figure*}

\subsection{Required power from neutrino observations}
The minimum cosmic-ray luminosity required from the IceCube neutrino data is shown in Fig.~\ref{fig:contour-decay}. We find that the cosmic-ray helium luminosity is $L_{\rm CR} \gtrsim 3\times 10^{46}~{\rm erg}~{\rm s}^{-1}$ for the whole parameter region.
The required cosmic-ray luminosity almost linearly increases with the emission radius, while being less sensitive to the spectral index.

Analytically, the differential muon neutrino luminosity after neutrino mixing is written as
\begin{eqnarray}
\varepsilon_\nu \frac{dL_{\varepsilon_{\nu_\mu}}}{d\varepsilon_\nu}
&\approx& \frac{2}{9}\frac{1}{2000} \varepsilon_n \frac{dL_{\varepsilon_n}}{d\varepsilon_n} \nonumber\\
&\approx & \frac{2}{9}\frac{1}{2000}\frac{1}{2}{\rm min}[1,f_{\rm dis}]\varepsilon_A \frac{dL_{\varepsilon_A}}{d\varepsilon_A},
\end{eqnarray}
where $\varepsilon_n (dL_{\varepsilon_n}/d\varepsilon_n)$ is the differential neutron luminosity. 
Here, we have considered that for the tribimaximal mixing of neutrinos from beta decay the flavor ratio on Earth becomes $\nu_e:\nu_\mu:\nu_\tau\approx5:2:2$.  
This implies, no matter what the effective optical depth is, the cosmic-ray power is larger than the muon neutrino luminosity by a factor of $\sim{10}^4$ and we have
\begin{equation}
L_{\rm CR} \gtrsim {10}^4 L_\nu \sim {10}^{46}~{\rm erg}~{\rm s}^{-1} \gg L_{\rm bol},  
\end{equation}
which agrees with our numerical simulations shown in Fig.~\ref{fig:contour-decay}. We also note that this conclusion about energetics constraints are insensitive to magnetic fields. 

\begin{figure}
    \centering    
    \includegraphics[width=0.5\textwidth]{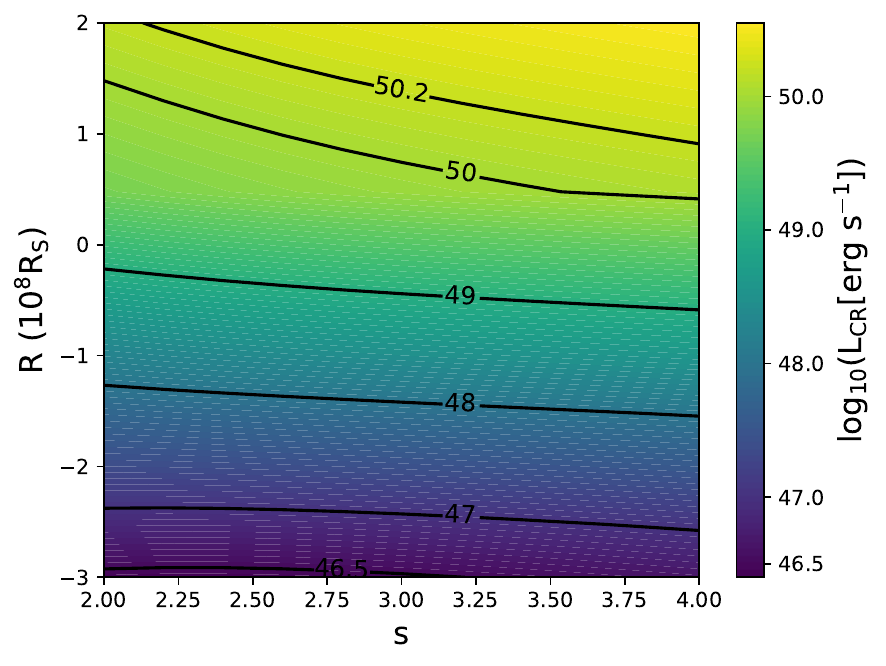}
    \caption{Required minimum cosmic-ray helium luminosity, $L_{\rm CR}$, as a function of $R$ and $s_{\rm CR}$. We find that $L_{\rm CR}$ exceeds $L_{\rm bol}\sim L_{\rm Edd}$ in all the parameter space.}
    \label{fig:contour-decay}
\end{figure}

\section{Discussions}\label{sec:dis}
\subsection{Implications for Coronal Environments}\label{sec:dis-imp}
We have found that larger values of $\xi_B$ more easily satisfy the \textit{Fermi}-LAT gamma-ray constraints because of synchrotron cascades~\citep{Murase:2022dog}. 
If the neutrino emission region coincides with the X-ray emission region, i.e., the hot corona, the ion plasma $\beta$ is 
\begin{equation}
\beta=\frac{8\pi n_p kT_p}{B^2}
%\approx\frac{\tau_T kT_p}{\zeta_\pm \sigma_T R U_\gamma}\xi_B^{-1} 
\approx \frac{\tau_T GM_{\rm BH}m_p}{\sqrt{3}\zeta_\pm \sigma_T R^2 U_\gamma} \xi_B^{-1}
\approx \left( \frac{\tau_T}{\sqrt{3}\zeta_e\lambda_{\rm Edd}}\right)\xi_B^{-1}
\end{equation}
where $T_p \approx GM_{\rm BH} m_p/(3 k R)$ is the virial temperature, $n_p \approx \tau_T /(\zeta_e\sigma_T H)$ is the nucleon density of the coronal plasma, $H$ is the coronal scale height, and
\begin{equation}
\frac{\tau_T GM_{\rm BH}m_p}{\sqrt{3}\zeta_e \sigma_T R^2 U_\gamma}\approx \frac{4\pi \tau_T GM_{\rm BH}m_pc}{\sqrt{3}\zeta_e \sigma_T L_{\rm bol}}=\frac{\tau_T}{\sqrt{3}\zeta_e\lambda_{\rm Edd}}, 
\end{equation}
which is $\sim0.1-1$ for ion-electron plasma because of $\tau_T\sim0.1-1$ for X-ray coronae, and $\sqrt{3}\lambda_{\rm Edd}\sim1$ for NGC 1068. Thus, high-$\beta$ plasma with $\beta\gtrsim 10$ means $\xi_B\lesssim0.01-0.1$, in which inverse-Compton cascades are expected, in which $R\lesssim3R_S$ should be satisfied. The synchrotron dominance for electromagnetic cascades occurs for $\xi_B\gtrsim0.1$~\citep{PhysRevLett.125.011101}, implying that $\beta\lesssim 1-10$ is necessary to allow $R\sim10R_S$. 

This favors the presence of low-$\beta$ plasma if hot coronae are responsible for high-energy neutrino emission. Together with our gamma-ray constraints on the emission radius, the most promising particle acceleration mechanisms would be magnetic dissipation, including magnetic reconnections \citep{Kheirandish:2021wkm,Mbarek:2023yeq,Fiorillo:2023dts} and associated stochastic acceleration in turbulence~\citep{PhysRevLett.125.011101,Kimura:2020thg, Eichmann:2022lxh} although some shock acceleration could be associated with magnetic reconnections~\citep{Murase:2022dog}. 

If the corona is not magnetically powered with high-$\beta$, the emission radius has to be even more compact. Although we do not exclude the entire parameter space, at least we exclude shock models with $R\sim(10-30)~R_S$ and weak magnetic fields, which are motivated by ALMA data at the submillimeter band~\citep{Inoue:2019yfs}. Note that our constraints are even more severe if the contribution of primary nonthermal electrons is included. 
On the other hand, we also note that this conclusion does not apply if the neutrino emission region is different from the hot corona responsible for X rays. For example, failed outflows \citep{AlvarezMuniz:2004uz,Inoue:2022yak} and the base of weak jets~\citep{Pe'er:2009rc,Murase:2022dog} may be viable. 

NGC 1068 is known to have a low-power jet with a luminosity of $L_j\sim{10}^{43}~{\rm erg}~{\rm s}^{-1}$, as indicated by radio observations of \emph{e-}Merlin and VLA~\citep{mutie2023radio} as well as ALMA~\citep{Michiyama_2022} through the empirical relationship~\citep{osti_21460066}. Knots within these jets may be responsible for gamma-ray emission in the range of $\sim0.1-1$~GeV~\citep{Salvatore:2023zmf}. The jet efficiency of jet-quiet AGNs is small, which is $L_j/L_{\rm bol}\sim0.01-0.1 \ll \lambda_{\rm Edd}$. It is believed that the efficiency of the jet launched via the Blandford-Znajek (BZ) mechanism~\citep{1977MNRAS.179..433B} for magnetically arrested disks is high. Based on this, \cite{inoue2024upper} speculated that coronae of Seyferts such as NGC 1068 have $\beta \gtrsim 10-100$. However, such high-$\beta$ coronae are strongly constrained by multimessenger observations, as shown in this work. Moreover, observationally, a relatively low jet efficiency of $\sim0.01-0.1$ has been inferred even for objects with powerful jets that are presumably launched by the BZ mechanism. The examples include Fanaroff-Riley II jets~\citep{Fan:2019ojg}, Fanaroff-Riley 0 jets~\citep{Khatiya:2023lkg}, and gamma-ray burst jets~\citep[e.g.,][]{Suwa_2011}, which could be caused by the time dependence of the magnetic flux and configurations. Low-$\beta$ coronae without powerful jets have also been inferred from numerical simulations~\citep[e.g.,][]{Jiang__2019}, and the accretion disk may also be strongly magnetized~\citep{hopkins2024forged,hopkins2024analytic}. 
Recent particle-in-cell \citep[e.g.,][]{Gro_elj_2024} and magnetohydrodynamic simulations \citep[e.g.,][]{10.1093/mnras/stad3261} have also supported that magnetic dissipation in magnetically-powered coronae is the most promising mechanism for electron heating.

\subsection{Particle Acceleration}\label{sec:dis-acc}
Our results support low-$\beta$ plasma in the coronal region or at the base of possible jets as an acceleration site, and disfavor accretion shocks at least for $R\gtrsim10R_S$. Among various possible acceleration processes, magnetic reconnections~\citep[e.g.,][]{Hoshino:2013pza, Hoshino:2015cka, Guo:2015ydj, 10.1093/mnras/stx2530, Ball:2018icx} and stochastic acceleration~\citep[e.g.,][]{Lynn:2014dya, Kimura:2016fjx, Comisso:2018kuh, Zhdankin:2018lhq, Kimura:2018clk, Lemoine:2020bsk, Wong:2019dog, Lemoine:2023wsw} are promising. 

The coronal plasma can be collisionless for ions although it may not be for electrons, in which stochastic acceleration may operate in the presence of magnetic turbulence. Whether plasma is low-$\beta$ or high-$\beta$, the timescale is given by 
\begin{align}
t_{\rm acc} &\approx \eta_{\rm tur} \left(\frac{c}{V_A}\right)^2 \frac{H}{c} \left(\frac{\varepsilon_p}{eBH}\right)^{2-q}\nonumber\\ 
&\simeq 1.8 \times 10^4 {\rm~s}~\eta_{\rm tur,1} \beta_{-1} 
{\left(\frac{R}{10 R_S}\right)}^{2} {\left(\frac{M_{\rm BH}}{10^7M_\odot}\right)} \nonumber \\ &\quad \left(\frac{\varepsilon_p}{eBH}\right)^{2-q},
\end{align}
where $\eta_{\rm tur}^{-1}$ is a parameter represents the turbulence strength, $q \sim 1.5-2$ describes the energy dependence of the diffusion coefficient, $V_A = B / \sqrt{4\pi m_p n_p}$ is the Alfv\'en velocity. 

For cosmic-ray protons with the energy around $100-1000$~TeV, the dominant cooling timescale is the Bethe-Heitler pair production cooling, which is
\begin{align}
    t_{\rm BH} &\approx \frac{1}{\tilde{n}_{\rm disk} \hat{\sigma}_{\rm BH} c} \nonumber \\ &\simeq 1.4 \times 10^3~{\rm~s}~\tilde{L}_{\rm disk, 44.7}^{-1} {\left(\frac{R}{10 R_S}\right)}^2 {\left(\frac{M_{\rm BH}}{10^7M_\odot}\right)}^2 \nonumber\\ &\quad\left(\frac{\varepsilon_{\rm disk}}{31.5\rm~eV}\right),
\end{align}
and the balance between acceleration and cooling time scales, $t_{\rm acc} \sim t_{\rm BH}$, may give $\varepsilon_{p}^{\rm max}\sim \mathcal{O}(100)$~TeV.

Particles may also be accelerated by magnetic reconnections. The magnetization parameter is 
\begin{eqnarray}
\sigma &=& \frac{B^2}{4\pi n_p m_p c^2} = \beta^{-1} \left(\frac{2k_B T_p}{m_pc^2}\right)= \beta^{-1} \left(\frac{R_S}{3 R}\right) \nonumber\\
&=& \frac{2 \xi_B \zeta_e L_{\rm bol} \sigma_T}{\sqrt{3} \tau_T m_p c^2 R}  \nonumber \\ 
&\simeq& 4.7\times 10^{-2} \frac{\xi_B \zeta_e}{\tau_T} L_{\rm bol, 45} \left(\frac{R}{10 R_S}\right)^{-1}  {\left(\frac{M_{\rm BH}} {10^7M_\odot}\right)}^{-1}.
% \nonumber \\ &\quad&
\end{eqnarray}
Particle acceleration via magnetic reconnections is efficient and lead to hard spectra for $\sigma>1$. For $R\lesssim 30R_S$, this requires $\beta\lesssim0.01-1$, so very low-$\beta$ plasma environments are necessary and extreme magnetization parameters such as $\sigma\gtrsim{10}^4$ have been considered~\citep{Mbarek:2023yeq,Fiorillo:2023dts}.    
The acceleration time scale is
\begin{align}
t_{\rm acc}&\approx \eta \frac{\varepsilon_p}{e B c} \nonumber \\ 
&\simeq 0.013~{\rm~s}~\eta_{1} \left(\frac{R}{10R_S}\right) 
\xi_B^{-1/2} L_{\rm bol, 45}^{-1/2} {\left(\frac{M_{\rm BH}}{10^7M_\odot}\right)} \nonumber \\
& \quad \left(\frac{\varepsilon_p}{100\rm~TeV}\right),
\end{align}
where $\eta$ is a coefficient describing the efficiency. At higher energies, the dominant cooling process becomes the photomeson production, whose timescale is given by
\begin{align}
t_{p\gamma} &\approx \frac{1}{\tilde{n}_{\rm disk} \hat{\sigma}_{\rm p\gamma} c} \nonumber \\ &\simeq 16~{\rm~s}~\tilde{L}_{\rm disk, 44.7}^{-1} {\left(\frac{R}{10 R_S}\right)}^2 \left(\frac{M_{\rm BH}}{10^7 M_\odot}\right)^{2} \nonumber\\
    &\quad\left(\frac{\varepsilon_{\rm disk}}{31.5\rm~eV}\right),
\end{align}
for cosmic-ray protons at sufficiently high energies.
The condition $t_{\rm acc}\sim t_{p\gamma}$ gives a maximum energy of $\varepsilon_{p}^{\rm max} \sim \mathcal{O}(10)$~\rm~PeV.

In the leptonic scenario, the acceleration time scale of electrons is 
\begin{eqnarray}
t_{\rm acc}\approx \eta \frac{\varepsilon_e}{e B c},
\end{eqnarray}
and the balance between the synchrotron and inverse-Compton cooling gives the electron maximum energy, 
\begin{eqnarray}
\varepsilon_{e}^{\rm max}&\approx& \left(\frac{6\pi e}{\sigma_T B \eta (1 + Y)}\right)^{1/2} m_e c^2 \nonumber\\
&\simeq& 0.64~{\rm~TeV}~[\eta (1 + Y)]^{-1/2} \left(\frac{R}{10R_S}\right)^{1/2} \xi_B^{-1/4} \nonumber \\ && L_{\rm bol, 45}^{-1/4},    
\end{eqnarray}
where $Y$ denotes inverse-Compton Y parameter. The maximum energy required for the leptonic scenario is realized only if the magnetic field is sufficiently weak. 

In the strongly magnetized plasma, the inverse-Compton emission is highly suppressed. Note that the maximum synchrotron energy is limited as
\begin{equation}
\varepsilon_{\rm syn\gamma}^{\rm max} \approx \frac{3heB}{4\pi m_e c} \left(\frac{\varepsilon_{e}^{\rm max}}{m_e c^2}\right)^2\simeq 240\rm~MeV [\eta (1 + Y)]^{-1},   
\end{equation}
which is the so-called synchrotron burnoff limit~\citep{1983MNRAS.205..593G, 2002PhRvD..66b3005A, 2018pgrb.book.....Z}, independent of the magnetic field strength. 
This energy is much below 10~TeV, being insufficient for the leptonic scenario.

\subsection{Impacts of other parameters}
In this work, we adopt $V=0.1 c$ as the default velocity of the system. We note that the velocity can be smaller, which affects effective optical depths through $t_{\rm dyn}$. For example, the infall velocity is estimated to be 
\begin{equation}
V_{\rm fall} = \alpha V_K \sim 0.01 c \alpha_{-1} (R / R_S)^{-1/2},
\end{equation}
where $\alpha$ is the viscous parameter in the accreting flow, $V_K = \sqrt{GM_{\rm BH}/R} =c/\sqrt{2 (R/R_S)}$ is the Keplerian velocity. The effective optical depth to photomeson production is larger with smaller velocities, in which the cosmic-ray luminosity requirement can be relaxed especially for large values of $R$. For the photohadronic scenario, in the limit that the system is calorimetric, the cosmic-ray luminosity is required to be $L_{\rm CR}\gtrsim 10^{43}~f_{\rm sup}^{-1}~{\rm erg}~{\rm s}^{-1}$, which is consistent with the conclusion by \citet{Murase:2022dog} although our constraint is tighter due to the suppression factor by the Bethe-Heitler process.  
Similarly, for the beta decay scenario, the efficiency is higher so that the power requirement is relaxed for large radii, but it cannot be below $\sim{10}^{46}~{\rm erg}~{\rm s}^{-1}$. 
On the other hand, the gamma-ray power requirement in the leptonic scenario is not affected because $\tau_{\gamma\gamma\rightarrow e^+e^-}\gg1$ is satisfied.  

Although we primarily focus on $p\gamma$ scenarios, one may consider $pp$ scenarios, where neutrinos are mostly produced via inelastic collisions between cosmic rays and plasma gas~\citep{PhysRevLett.125.011101,Inoue:2019yfs,Eichmann:2022lxh}. The effective optical depth for inelastic $pp$ cooling rate is estimated to be
\begin{eqnarray}
    f_{pp} &\approx& n_p \hat{\sigma}_{pp} R (c / V) \nonumber \\ &\simeq& 0.26~(\tau_T / 0.5) \left(\frac{V}{0.1 c}\right)^{-1},
\end{eqnarray}
where $\hat{\sigma}_{pp} \sim 2 \times 10^{-26}\rm~cm^2$ is the effective $pp$ cross section including the inelasticity, $n_p \approx \tau_T / (\zeta_e \sigma_T H)$ is the nucleon density in the corona region, and $\tau_T \sim 0.1 - 1$.
Based on the comparison with Eq.~\ref{eq:pg-loss}, we see that the  $pp$ process can be more important than the $p\gamma$ process at large emission radii. However, our conclusions for $\xi_B\lesssim0.1$ would remain the same because cascades from photohadronic processes are more relevant at small emission radii, while $R=30R_S$ is allowed for $\xi_B\sim1$~\citep{Murase:2022dog,Ajello:2023hkh}.  

Finally, we comment on the flavor ratio. In the standard hadronic scenarios, the approximate flavor ratio observed on Earth is expected to be $\nu_e:\nu_\mu:\nu_\tau \approx 1:1:1$, which is assumed in the IceCube data shown in Fig.~\ref{fig:spectrum-plots}. 
However, the flavor ratio varies among different mechanisms, and we have
\begin{eqnarray}
\nu_e:\nu_\mu:\nu_\tau \approx 1:1:1  \,\,\,\,\,  ({\rm hadronic})\nonumber \\
\nu_e:\nu_\mu:\nu_\tau \approx 14:11:11 \,\,\,\,\, ({\rm leptonic})\nonumber \\
\nu_e:\nu_\mu:\nu_\tau \approx 5:2:2  \,\,\,\,\,  ({\rm beta~decay})
\end{eqnarray}
Correspondingly, in Figs.~\ref{fig:result-lepto} and \ref{fig:result-decay}, different flavor ratios are assumed when the IceCube data are depicted. The determination of neutrino flavors with future observations will also be relevant as an independent method to discriminate among different neutrino production mechanisms~\citep[e.g.,][]{Beacom:2003nh,Shoemaker:2015qul,Bustamante:2019sdb}. 
The IceCube Collaboration has analysed the flavor composition using all-sky neutrino data~\citep{Aartsen:2015ivb,Aartsen:2015ita,lad2023summary}. It has been proposed that AGNs like NGC 1068 can account for the all-sky neutrino flux in the $10-100$~TeV range~\citep{PhysRevLett.125.011101}, and the all-sky neutrino data can also be used to discriminate among different production mechanisms. In particular, the beta decay scenario has already been ruled out as the dominant neutrino production mechanism~\citep{Bustamante:2019sdb}.

\section{Summary}\label{sec:summary}
In this study, based on the recent multimessenger observations including IceCube and {\it Fermi}-LAT data, we constrained the cosmic-ray luminosity and emission radius that are compatible with Seyfert II galaxy NGC 1068. 
Our results are summarized as follows. 

\begin{enumerate}
    \renewcommand{\labelenumi}{(\roman{enumi})}
    \item In the photohadronic scenario, the emission radius is constrained to $R\lesssim(3-15)~R_S$. These results support the previous findings by \citet{Murase:2022dog} but thanks to the updated {\it Fermi}-LAT data~\citep{Ajello:2023hkh}, our constraints are stronger. For these emission radii, the required minimum cosmic-ray proton luminosity in the energy range of $10-30$~TeV is $L_{\rm CR}\gtrsim (3-10)\times 10^{43}$~erg~s$^{-1}$ \citep[see also Section~2.3 of][]{Murase:2022dog}. The minimum cosmic-ray luminosity can be lower than the total X-ray luminosity, which is also consistent with considerations from the all-sky neutrino and gamma-ray flux data~\citep{Murase:2015xka}.  
       
    \item In the leptonic scenario, neutrinos come from the muon-antimuon pair production~\citep{Hooper:2023ssc}. We found that cascaded gamma rays always overshoot gamma-ray and/or X-ray data. The muon pair production efficiency is too low when coronal X-ray emission is considered as a target photon field. This conclusion remains unchanged when the effects of nonlinear electromagnetic cascades are taken into account, which makes our conclusion more robust. We showed that the minimum gamma-ray luminosity is $L_{\gamma}\sim 10^{47}$~erg~s$^{-1}$ both analytically and numerically, which largely exceeds the Eddington luminosity.  
    We further argued that it is challenging to accelerate electrons up to 10~TeV energies in strongly magnetized plasma. 
    From the above three general arguments, we conclude that the leptonic scenario is unlikely as a viable mechanism for neutrino emission from NGC 1068. 

    \item In the beta decay scenario, neutrinos mainly originate from neutrons generated via the photodisintegration of nuclei~\citep{Yasuda:2024fvc}. With {\sc AMES}, which quantitatively handles details on the spectrum of neutrons from the disintegration and subsequent beta decay spectra, we found that inverse-Compton emission by beta decay electrons violates the gamma-ray data when observed infrared photons from the dust torus are considered. This conclusion is robust when the inevitable contribution from the Bethe-Heitler pair production is considered, and even stronger when the photomeson production is consistently included. We also showed that the minimum cosmic-ray power to explain the neutrino data in this scenario, $L_{\rm CR}\gtrsim 10^{46}$~erg~s$^{-1}$, largely exceeds the Eddington luminosity, by which we concluded that the beta decay scenario is excluded.  

    \item Our results support the standard neutrino production mechanisms that high-energy neutrinos are produced by $pp$ or $p\gamma$ processes, which require compact emission radii~\citep{Murase:2022dog}, and rule out the leptonic and neutron-decay scenarios without depending on any details of the models.  
    Our conclusions rely on two different general arguments: X-ray/gamma-ray constraints and cosmic-ray energetics required by neutrino observations. In particular, the former is robust because the ratio of electromagnetic to neutrino components is relevant without depending on the values of optical depths such as $\tau_{\gamma \gamma \to \mu^+ \mu^-}$ and $f_{\rm dis}$ themselves. 
    In the photohadronic scenario, we found that the gamma-ray constraints on $R$ depend on $\xi_B$, depending on whether cascades primarily occur via synchrotron or inverse-Compton processes. For $\xi_B\lesssim0.1$ corresponding to $\beta\gtrsim10$, electromagnetic cascades occur via inverse-Compton emission, in which the {\it Fermi}-LAT data led to $R\lesssim 3 R_S$. This disfavors high-$\beta$ plasma regions as particle acceleration sites, and accretion shock models with $R\gtrsim(10-30)R_S$~\citep[e.g.,][]{Inoue:2019yfs} are unlikely. 
    On the other hand, for $\xi_B\gtrsim 1$ corresponding to $\beta\lesssim0.1$, the gamma-ray constraint on the emission region is relaxed to $R\lesssim15~R_S$, which is consistent with the magnetically-powered corona model~\citep{PhysRevLett.125.011101}. 

\end{enumerate} 

We showed that the main production mechanism for high-energy neutrinos from NGC 1068 must be either the photohadronic ($p\gamma$) or hadronuclear ($pp$) mechanism. 
Our results demonstrated how multimessenger modeling of neutrino and gamma-ray data can be used for revealing and constraining neutrino production mechanisms. 
We also obtained some implications for possible acceleration mechanisms through cosmic-ray energetics constraints and constraints on neutrino emission regions. The calculations presented in this work are applicable to not only NGC 1068 but also other AGNs such as NGC 4151 and NGC 4945~\citep{Murase:2023ccp}.
As discussed in this work, measurements of the neutrino flavor ratios are also useful as a complementary way to confirming our conclusion.

%%%%%%%%%%%%%%%%%%%%%%%%%%%%%%%%%%%%%%%%%%%%%%%%%%
%%%%%%%%%%%%%%%%%%%%%%%%%%%%%%%%%%%%%%%%%%%%%%%%%%

\begin{acknowledgements}
First of all, we thank the reviewer for his valuable comments. We thank Mukul Bhattacharya, Carlos Blanco, and Mainak Mukhopadhyay for the useful discussions. 
The research by A.D. and K.M. is supported by the NSF Grant No.~AST-2308021. The work of K.M. is supported by the NSF Grants No.~AST-2108466 and No.~AST-2108467. The research by B.T.Z. and K.M. is supported by KAKENHI No.~20H05852.
\end{acknowledgements}

%\appendix

\bibliography{kmurase}
\bibliographystyle{aasjournal}

\end{document}